# Strong dust processing in circumstellar discs around 6 RV Tauri stars★,★★

## Are dusty RV Tauri stars all binaries?

S. De Ruyter[1], H. Van Winckel[2], C. Dominik[3], L. B. F. M Waters[2,3], and H. Dejonghe[1]

[1] Sterrenkundig Observatorium, Universiteit Gent, Krijgslaan 281 S9, 9000 Gent, Belgium
 e-mail: stephanie.deruyter@UGent.be
[2] Instituut voor Sterrenkunde, KU Leuven, Celestijnenlaan 200B, 3001 Leuven, Belgium
[3] Sterrenkundig Instituut "Anton Pannekoek", Kruislaan 403, 1098 Amsterdam, The Netherlands

**Abstract.** We present extended Spectral Energy Distributions (SEDs) of seven classical RV Tauri stars, using newly obtained submillimetre continuum measurements and Geneva optical photometry supplemented with literature data. The broad-band SEDs show a large IR excess with a black-body slope at long wavelengths in six of the seven stars, R Sct being the noticeable exception. This long wavelength slope is best explained assuming the presence of a dust component of large grains in the circumstellar material. We show that the most likely distribution of the circumstellar dust around the six systems is that the dust resides in a disc. Moreover, very small outflow velocities are needed to explain the presence of dust near the sublimation temperature and we speculate that the discs are Keplerian. The structure and evolution of these compact discs are as yet not understood but a likely prerequisite for their formation is that the dusty RV Tauri stars are binaries.

## 1. Introduction

RV Tauri stars form a distinct but loosely defined class of classical pulsating objects. They are located in the brightest part of the Population II instability strip and are defined as luminous (I-II) mid-F to K supergiants that show alternating deep and shallow minima in their light curves. They have formal periods (defined as the time between two successive deep minima) between 30 and 150 days, but often show strong cycle-to-cycle variability. RV Tauri stars are subdivided into two photometric classes on the basis of their light curves. The RVa class contains the stars with constant mean magnitude, while the RVb stars show variation of the mean magnitude on a timescale in the range of 600 to 1500 days.

After Gehrz (1972) and Lloyd Evans (1985) pointed out that many RV Tauri stars present near-infrared excesses, Jura (1986) identified the low-mass RV Tauri variables as young post-AGB objects. IRAS detected in many RV Tauri stars considerable amounts of cool, circumstellar dust, which is interpreted as being a relic of the strong dusty mass-loss on the AGB.

Chemically, however, RV Tauri stars do *not* show the expected post-AGB abundances: no high C-abundances or s-process overabundances are observed. Instead, RV Tauri photospheres often show chemical anomalies similar to the depletion patterns seen in the gas phase of the ISM (e.g. Giridhar et al. 2000). These patterns are a result of a poorly understood chemical process in which the separation of the circumstellar dust from the circumstellar gas is followed by a selective reaccretion of only the gas, which is then rich in non-refractory elements. Waters et al. (1992) proposed that the most likely circumstance for the process to occur is when the dust is trapped in a circumstellar disc. Note, however, that there is to date no detailed quantitative model of this reaccretion scenario.

Originally, depletion patterns were found in $\lambda$ Boo stars and only in five peculiar post-AGB stars, which were indeed proven to be binaries likely surrounded by a stable circumbinary disc (Van Winckel et al. 1995). The depletion patterns observed in many RV Tauri stars might call for the presence of such a disc as well, but since there is no relation found between the observed IR excess and the chemical composition of the stellar photosphere, there is a priori no direct observational evidence for this claim (e.g. Giridhar et al. 2000, and references therein).

**Table 1.** The name, the HD number, the equatorial coordinates $\alpha$ and $\delta$ (J2000), the Galactic coordinates $l$ and $b$, the formal period $P$, the amplitude in the $V$-magnitude $\Delta V$, the photometric classification, the effective temperature $T_{\rm eff}$, the surface gravity $\log g$ and the metallicity [Fe/H] of the seven classical RV Tauri stars from our sample.

| Name   | HD number | $\alpha$ (J2000) (h m s) | $\delta$ (J2000) (° ′ ″) | $l$ (°) | $b$ (°) | Period (days) | $\Delta V$ (mag) | $T_{\rm eff}$ (K) | $\log g$ (cgs) | [Fe/H] (dex) | Class |
|--------|-----------|--------------------------|--------------------------|---------|---------|---------------|------------------|-------------------|----------------|--------------|-------|
| TW Cam |           | 04 20 47.63              | +57 26 28.78             | 148.26  | +5.26   | 85.6          | 1.10             | 4800[1]           | 0.0[1]         | −0.5[1]      | RVa[5] |
| RV Tau | HD 283868 | 04 47 06.74              | +26 10 45.2              | 174.77  | −12.19  | 78.698        | 3.50             | 4500[1]           | 0.0[1]         | −0.4[1]      | RVb[5] |
| SU Gem | HD 42806  | 06 14 00.02              | +27 42 12.3              | 184.18  | +4.81   | 50.12         | 2.30             | 5750[3]           | 1.125[3]       | −0.7[3]      | RVb[5] |
| UY CMa |           | 06 18 16.37              | −17 02 34.81             | 224.75  | −14.85  | 113.9         | 2.00             | 5500[4]           | 1[4]           | +0.0[4]      | RVa[5] |
| U Mon  | HD 59693  | 07 30 47.47              | −09 46 36.80             | 226.14  | +4.15   | 92.26         | 2.00             | 5000[1]           | 0.0[1]         | −0.8[1]      | RVb[5] |
| AC Her | HD 170756 | 18 30 16.24              | +21 52 00.62             | 50.49   | +14.24  | 75.4619       | 2.31             | 5500[2]           | 0.5[2]         | −1.5[2]      | RVa[5] |
| R Sct  | HD 173819 | 18 47 28.95              | −05 42 18.6              | 27.40   | −1.72   | 140.2         | 1.99             | 4500[1]           | 0.0[1]         | −0.4[1]      | RVb[5] |

References: 1. Giridhar et al. (2000); 2. Van Winckel et al. (1998); 3. Wahlgren (1992); 4. model parameters estimated on the basis of the spectral type; 5. Lloyd Evans (1985).

However, for individual objects like AC Her, the presence of a stable circumstellar structure has been proposed to explain the ISO SWS spectrum (Van Winckel et al. 1998) and the CO data (Jura & Kahane 1999). The disc was even claimed to be resolved in the $N$ and $Q$-band (Jura et al. 2000) but this was not confirmed by higher resolution imaging (Close et al. 2003). Indirect observational evidence for the existence of a disc in more objects includes the presence of a near-IR excess in the SED (Lloyd Evans 1999). Also CO ($J = 1-0$) rotational line emission is remarkably weak and narrow for the few objects detected (Bujarrabal et al. 1988), indicating either very slow winds or Keplerian kinematics.

The dusty particles in a disc may grow by coagulation to sizes much larger than found in AGB outflows, and one way to identify those large particles isto observe at long wavelengths. We have therefore obtained submillimetre continuum observations at 850 $\mu$m (Sect. 2) with the Submillimetre Common-User Bolometer Array (SCUBA) at the James Clerk Maxwell Telescope (JCMT) of 7 classical RV Tauri stars (Table 1).

In Sect. 3 the SEDs of the 7 classical RV Tauri stars are presented from the UV up to submillimetre wavelengths. Although the objects are well-known RV Tauri stars, we realized that the total line-of-sight extinction of the sample was very poorly constrained in the literature. We therefore quantify in a systematic way the total extinction of all objects based on broad-band photometry. As a first-order approximation, an optically thin dust model is used to fit the observed IR excess in Sect. 4. The results are evaluated with respect to the dust geometry in Sect. 5 and in Sect. 6 conclusions are formulated.

## 2. Observations and data reduction: 850 $\mu$m SCUBA data

Observations were carried out with the 15 m JCMT at Mauna Kea, Hawaii, during March-April 1999 (programme M99AN08). SCUBA (Holland et al. 1999) was operated in photometry mode to simultaneously obtain data at 850 $\mu$m and 450 $\mu$m. For the data reduction, the standard software SURF was used with Mars as a flux calibrator. The 850 $\mu$m contin-uum measurements are shown in Table 2, while for 450 $\mu$m, no significant signal was detected.

**Table 2.** Observations with SCUBA: 850 $\mu$m fluxes.

| Name   | $F_{850}$ (mJy) |
|--------|-----------------|
| TW Cam | 11.0 ± 2.3      |
| RV Tau | 50.3 ± 3.6      |
| SU Gem | 7.5 ± 2.5       |
| UY CMa | 2.4 ± 2.1       |
| U Mon  | 181.6 ± 2.6     |
| AC Her | 99.4 ± 3.8      |
| R Sct  | 2.4 ± 1.6       |

## 3. Spectral Energy Distributions (SEDs)

### 3.1. (Spectro-)Photometric data

The main difficulty in constructing the SEDs of pulsating stars with large amplitudes and in many cases severe cycle-to-cycle variability is the acquisition of equally phased data over a wide spectral domain. Since these data are not available, we limited our study of the broad-band energetics to the phase of minimal and maximal brightness.

We acquired Geneva optical photometry (Table A.1) at random epochs for RV Tau (53 measurements), SU Gem (29), U Mon (119), AC Her (62) and R Sct (36) with the Flemish Mercator Telescope at La Palma, using the refurbished Geneva photometer P7 (Raskin et al. 2004). We added additional Geneva optical photometry from the Geneva database (http://obswww.unige.ch/gcpd/gcpd.html). Johnson and Cousins broad-band photometry was found in the literature (Table A.2). For TW Cam and UY CMa no Geneva photometry is available, so we used only literature data. Near-IR data were taken from the 2 MASS and DENIS projects complemented with data from the literature (Table A.3). Far-IR data come mainly from IRAS (Table A.4) while for SU Gem, U Mon and R Sct, MSX data are also available (Table A.5). At the short wavelength side we used the IUE (0.115 $\mu$m–0.320 $\mu$m) data from the newly extracted spectral data release (INES). The resulting SEDs are given in Figs. 1 and 2.

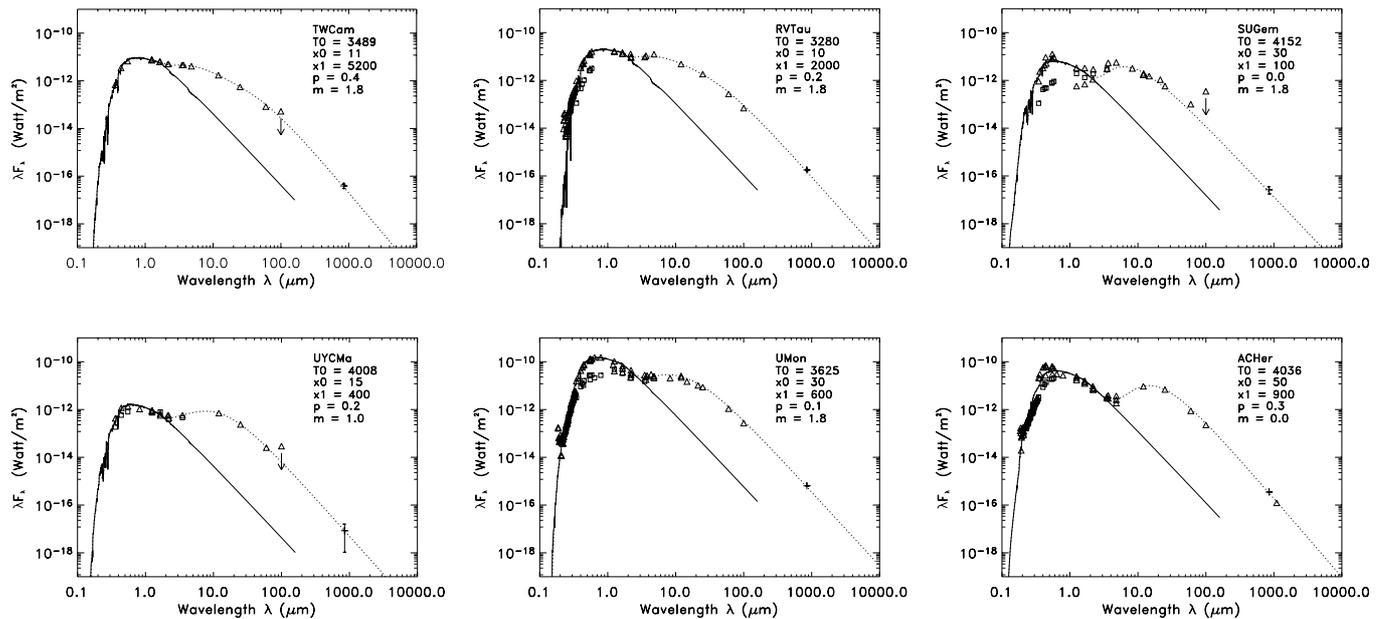

**Fig. 1.** The SEDs of 6 RV Tauri stars: the dereddened fluxes are given together with the scaled photospheric Kurucz model representing the unattenuated stellar photosphere (solid line). An optically thin dust fit was used to model the IR excess (dotted line). Data found in the literature together with our 7 band Geneva photometry (only the maxima) are plotted as triangles. The minimal data points (squares) were not used for the determination of $E(B-V)$. They are only plotted to give an idea of the amplitude of the pulsations. Crosses represent our 850 $\mu$m SCUBA data point. Error bars on the 850 $\mu$m SCUBA data point are plotted as well, though for some objects smaller than the symbols. The arrow at the 100 $\mu$m fluxpoint shows that this data point is an upper limit for some stars.

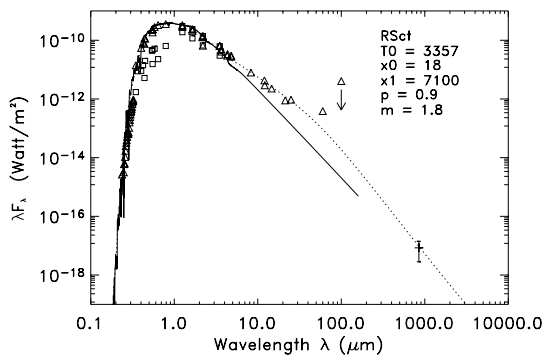

**Fig. 2.** The SED of R Sct is clearly different from these of the other programme stars. Symbols are the same as in Fig. 1.

### 3.2. Colour excess $E(B-V)$ determination

To determine the total line-of-sight reddening, we used only the most luminous data. The main reason is that for RVb stars in long-term minimum, the circumstellar extinction is likely maximal and often even dominates the interstellar component. Moreover, the effective temperature determinations are based on the analysis of high-resolution spectra used in the chemical analyses of the stars and those spectra are preferentially taken at maximum light, because of the minimal molecular veiling during the hotter phase in the light curve (Giridhar et al. 2000).

Changing the stellar models by $\pm 250$ K in effective temperature implies an uncertainty in $E(B-V)$ of about 0.2.

Because of the circumstellar material, the total extinction must contain both a circumstellar and an interstellar component. Whether the circumstellar extinction affects the line-of-sight depends on the geometry. The estimate for $E(B-V)$ is found by minimizing the difference between the dereddened observed fluxes in the UV-optical and the appropriate Kurucz model (Kurucz 1979). We scale our data to the $J$ filter, the reddest filter for which no dust excess is expected. For the dereddening of the observed maximal fluxes we used the average interstellar extinction law given by Savage & Mathis (1979). This implies that we also adopt the interstellar extinction law for the circumstellar component of the total extinction.

The total colour excess for all objects lies between 0.0 and 0.8 (Table 4). The intrinsic error on this $E(B-V)$-determination is typically 0.1 . Together with the error of 0.2 induced by the uncertainty of the temperature of the underlying photosphere, we have an uncertainty of 0.3 on the total extinction during maximal light.

Despite the significant IR excess, the total line-of-sight extinction is very small for UY CMa, U Mon and R Sct, even within the error. While for R Sct the dust excess is marginal and the large amplitude pulsations make the determination of the total reddening uncertain, the very large IR excess of UY CMa and U Mon is clearly in contradiction with the lack of line-of-sight reddening. This implies either a significant grey extinction or a non-spherically symmetric circumstellar dust distribution.

**Table 3.** The five parameters of the dust model: the normalization temperature $T_0$, the inner radius of the dust shell $r_{in}$, the outer radius of the dust shell $r_{out}$, the spectral index $p$ and the density parameter $m$. The last column lists the energy ratio $L_{IR}/L_*$.

| Name | $T_0$ (K) | $r_{in}$ ($R_*$) | $r_{in}$ (AU) | $r_{out}$ ($R_*$) | $r_{out}$ (AU) | $p$ | $m$ | $L_{IR}/L_*$ |
|---|---|---|---|---|---|---|---|---|
| TW Cam | 3489 | 11 | 4.5 ± 1.6 | 5200 | 2140 ± 890 | 0.4 | 1.8 | 0.45 ± 0.06 |
| RV Tau | 3280 | 10 | 4.7 ± 1.7 | 2000 | 930 ± 340 | 0.2 | 1.8 | 0.43 ± 0.06 |
| SU Gem | 4152 | 30 | 4.9 ± 1.6 | 100 | 16.3 ± 5.4 | 0.0 | 1.8 | 0.34 ± 0.05 |
| UY CMa | 4008 | 15 | 5.3 ± 2.0 | 400 | 138 ± 52 | 0.2 | 1.0 | 0.37 ± 0.05 |
| U Mon | 3625 | 30 | 11.5 ± 4.2 | 600 | 229 ± 84 | 0.1 | 1.8 | 0.07 ± 0.01 |
| AC Her | 4036 | 50 | 12.5 ± 4.4 | 900 | 224 ± 80 | 0.3 | 0.0 | 0.014 ± 0.002 |
| R Sct | 3357 | 18 | 13.4 ± 5.3 | 7100 | 5300 ± 2100 | 0.9 | 1.8 | 0.020 ± 0.003 |

**Table 4.** The estimated total colour excess $E(B-V)$ and the optical depth estimates in the V-band. $\tau_{V,E(B-V)}$ is determined from the SED minimalization and gives the total line-of-sight extinction. The contribution from the ISM is given in Col. 3 assuming the distance range given in Table 5. $\tau_{V,\text{dust model}}$ was determined by integrating the spherically symmetric dust model and in the last column the mean optical depth is given, determined by the energy ratio from the dust excess over the optical fluxes.

| Name | $E(B-V)$ | $\tau_V$ | | | |
|---|---|---|---|---|---|
| | | $E(B-V)$ | ISM | model | mean |
| TW Cam | 0.3 ± 0.3 | 0.86 | $2.68_{2.10}^{3.27}$ | 0.62 | 0.60 |
| RV Tau | 0.4 ± 0.3 | 1.14 | $1.75_{1.75}^{1.75}$ | 0.61 | 0.56 |
| SU Gem | 0.8 ± 0.3 | 2.28 | $1.04_{1.04}^{1.47}$ | 0.55 | 0.42 |
| UY CMa | 0.0 ± 0.3 | 0.00 | $0.00_{0.00}^{0.00}$ | 0.75 | 0.46 |
| U Mon | 0.3 ± 0.3 | 0.00 | $0.51_{0.35}^{0.56}$ | 0.22 | 0.07 |
| AC Her | 0.3 ± 0.3 | 0.86 | $0.62_{0.48}^{0.63}$ | 0.27 | 0.01 |
| R Sct | 0.0 ± 0.3 | 0.00 | $1.28_{0.98}^{1.51}$ | 0.03 | 0.02 |

### 3.3. Luminosity and distance estimates

The intrinsic luminosities of our sample stars were estimated using the extension of the type II Cepheid period-luminosity relation based on RV Tauri stars in the Large Magellanic Cloud (Alcock et al. 1998). For variables with $P/2 > 12.6$ days we use

$$M_V = 2.54\,(\pm 0.48) - (3.91\,(\pm 0.36))\log(P/2), \sigma = 0.35.$$

The results are given in Table 5, together with the errors. The objects are luminous but the large true intrinsic scatter in the $P - L$ relation prevents a more accurate estimate. The dereddened and scaled model atmosphere yield the distance estimates presented in Table 5. Only for 2 programme stars were parallaxes obtained by Hipparcos. For U Mon the parallax 1.45 ± 0.82 mass gives the distance range 440−690−1587 pc compatible with the $P - L$ distance, while for R Sct the parallax 2.32 ± 0.82 mass gives a 318−431−667 pc which is clearly on the short side of the $P - L$ distance.

## 4. An optically thin dust fit

A broad-band SED is limited in diagnostic value for constraining the chemistry and spatial distribution of the circumstellar dust. The line-of-sight reddening of the systems is not large, so as a simple first-order approximation, we apply a spherically symmetric, optically thin dust model described by Sopka et al. (1985). In this model, the flux $F_\nu$ is determined by five parameters: the normalization temperature $T_0$, the inner radius of the dust shell $r_{in}$, the outer radius of the dust shell $r_{out}$, the spectral index $p$ and the density parameter $m$. We apply a least square minimalization to determine the free parameters $r_{in}$, $r_{out}$, $p$ and $m$. Results are given in Table 3.

**Table 5.** The bolometric corrections $BC_V$ (from Bessell et al. 1998), the luminosity $L$ (in $10^3\,L_\odot$), the distance $D$ (in kpc) and the dust mass ($10^{-5}\,M_\odot$) are given, together with the estimated errors.

| Name | $BC_V$ | $L$ ($10^3\,L_\odot$) | $D$ (kpc) | $M_d$ ($10^{-5}\,M_\odot$) |
|---|---|---|---|---|
| TW Cam | −0.35 | 3.7 ± 2.6 | 3.1 ± 1.1 | 3.0 ± 2.2 |
| RV Tau | −0.49 | 3.7 ± 2.6 | 2.17 ± 0.72 | 4.2 ± 3.0 |
| SU Gem | −0.03 | 1.20 ± 0.77 | 2.11 ± 0.66 | 0.45 ± 0.32 |
| UY CMa | −0.08 | 4.5 ± 3.3 | 8.4 ± 3.1 | 3.3 ± 3.8 |
| U Mon | −0.24 | 3.8 ± 2.7 | 0.77 ± 0.28 | 1.8 ± 1.3 |
| AC Her | −0.07 | 2.4 ± 1.6 | 1.13 ± 0.39 | 3.0 ± 2.1 |
| R Sct | −0.52 | 9.4 ± 7.1 | 0.75 ± 0.29 | 0.10 ± 0.10 |

We find small spectral indices in the range from 0.0 to 0.4 for all objects, with the noticeable exception of R Sct. In the SEDs, the 850 $\mu$m fluxes indeed follow a Rayleigh-Jeans slope from the 60−100 $\mu$m fluxpoint redwards (Fig. 1). Such slopes are consistent with the presence of a composition of large grains (radius $\gtrsim 0.1$ mm) in the circumstellar environment of the RV Tauri stars. For R Sct, however, we find $p \approx 1$. R Sct is a rather unusual RV Tauri and Matsuura et al. (2002) show that its period evolution is slower than expected, implying that R Sct is presently not evolving towards higher temperatures, which is in contrast with the post-AGB tracks. They argue that R Sct may be a thermal-pulsing AGB star, observed in a helium-burning phase instead of a post-AGB star.

An important conclusion of our dust fitting is also that the inner radii $r_{in}$ are very small: in all cases less than 14 AU from the central object. The dust is located very close to the star and dust excesses start near the sublimation temperature. The distance of the inner boundary to the stellar surface would be covered by a normal AGB wind velocity of 10 km s$^{-1}$ in only about 7 years. If indeed the dust we observe is a relic of the AGB mass-loss, the expansion velocity of the circumstellar material must be extremely small! Note that there is no evidence of current day dusty mass-loss. Moreover, RV Tauri stars with very similar pulsational amplitudes and spectral types exist without detected infrared excesses.

Although the outer radius estimates are not well constrained and are very dependent on the exact choice of the spectral index, it is clear that the outer radii are not very large either. In most cases, the dust can be fitted well, even with a single temperature black-body (e.g. Dominik et al. 2003). This is also clear from visual inspection of the SED plots, which show a lack of cool ($T_d \leq 100$ K) dust. If the circumstellar geometry is optically thick, the extent of the dust distribution would be even more compact than in the optically thin case. Given the distances to the objects, interferometric observations will be needed to resolve the circumstellar dust geometry.

## 5. Discussion: Characteristics of the dust excess

### 5.1. Dust geometry

To test whether spherical symmetry of the circumstellar dust is a reasonable assumption, we investigate the balance between our line-of-sight extinction and the visual extinction computed by integrating our dust model. The ISM contribution of the total extinction was determined using the Galactic extinction routine from Hakkila et al. (1997) which averages several three-dimensional Galactic extinction models. Unfortunately, the real diagnostic value of this part of the study is somewhat limited because of the significant uncertainties involved. A spatially more detailed interstellar line-of-sight determination towards the RV Tauri objects would be helpful.

Nevertheless, the different optical depths in Table 4 yield significant information on individual objects. At first sight, only the total reddening of AC Her is the sum of the ISM and the circumstellar component. UY CMa, U Mon and RV Tau are examples for which the large IR excess is inconsistent with the small total line-of-sight reddening, pointing to a strongly non-spherical geometry of the circumstellar material. For SU Gem the total reddening is larger than the sum of the likely ISM contribution and the circumstellar extinction expected from a spherical dust shell. For TW Cam, the large expected ISM contribution indicates that our distance estimates are too large. The empirical extinction model from Neckel et al. (1980) shows a steep rise up to about 1 kpc in the direction of TW Cam. Again also in this star the circumstellar line-of-sight extinction must be very small.

R Sct is a case for which the period-luminosity relation does not hold. The Galactic coordinates imply a significant mean interstellar component of the extinction using the distance obtained with the $P - L$ relation (Table 4). Even with the upper limit of the Hipparcos parallax, one expects the ISM extinction to be significant ($E(B-V)_{ISM} = 0.7$ using the mean exinction estimates from Hakkila et al. (1997) with a distance of 320 pc). R Sct is known to be a very irregular pulsator with a large amplitude, but strict simultaneous photometric studies over a wide wavelength domain show a small total extinction (e.g. $E(B-V) = 0.3-0.5$, Shenton et al. 1994) for which the interstellar component is low (Cardelli 1985, $E(B-V)_{ISM} = 0.2$). We conclude that R Sct is much less luminous than given in Table 5. This RV Tauri object is, in line with other studies (e.g. Matsuura et al. 2002), considered as an exception.

Our energy balance calculations indicate a non-spherical distribution of the circumstellar dust in 5 sources, with the possible exception of SU Gem, where we are likely looking into the disc.

The amount of energy reprocessed by the dust grains ($L_{IR}/L_*$) is larger than 34% for all objects, except for U Mon, AC Her and R Sct (Table 3). The mean optical depth of the circumstellar dust (Table 4) is defined as $\tau_{mean} = -\ln(1-(L_{IR}/L_*))$. Assuming that the IR emission is produced by an infinitely optically thick disc, the opening angle of the disc as seen from the star must be $\theta \sim 44°$ for objects with 40% flux conversion. The significant scaleheight is likely to be sustained by gas pressure in the gas-rich discs (e.g. Dominik et al. 2003).

Despite the large scaleheights, we still prefer to use the word "disc" instead of torus. Dust tori are often resolved around Proto Planetary Nebulae, but these have very different SED characteristics since they are much colder and are likely expanding (e.g. Sánchez Contreras & Sahai 2004). Moreover, the physical sizes of the resolved tori are much larger than we expect the circumstellar material to be in the RV Tauri stars.

### 5.2. Dust mass $M_d$

The 850-$\mu$m fluxes can be used to estimate the mass of the dust as the dust is likely to be optically thin at this wavelength. Using the Rayleigh-Jeans form for the Planck function and assuming that the dust is at a mean temperature $T_d = 500$ K, the dust mass is given by (e.g. Hildebrand 1983)

$$M_d (M_\odot) = 2.5052 \times 10^{-6} \frac{F_{850} \text{ (mJy)} \, D^2 \text{ (kpc)}}{\kappa_{850} \text{ (cm}^2 \text{ g}^{-1})}.$$

With the assumption that $\lambda \gg a$, we used the values $Q_0 = 7.5 \times 10^{-4}$ and $\lambda_0 = 125 \, \mu$m (for $a = 0.1 \, \mu$m) given by Hildebrand (1983) in the power law approximation $Q_\nu = Q_0 \, (\lambda_0/\lambda)^p$ of the emission efficiency factor. With $\rho_{gr} = 3$ g cm$^{-3}$ and using our calculated values for $p$, we obtain dust mass absorption coefficients $\kappa_{850}$ for each source.

In each case, the dust mass lies within the range of $1 \times 10^{-5} \, M_\odot$ and $5 \times 10^{-5} \, M_\odot$, with the noticeable exception of R Sct for which $1 \times 10^{-6} \, M_\odot$ is found (Table 5).

## 6. Conclusions

We presented the extended SEDs of 7 classical RV Tauri stars based on literature data, supplemented with our own Geneva photometry and new continuum measurements at 850 $\mu$m. The objects were selected to have a large infrared excess as detected by IRAS. Our main conclusions can be summarized as follows:

– Our systematic determination of the total extinction implies large energy conversion ratios $L_{IR}/L_*$, up to 45%.

- The SEDs reveal the presence of a very hot dust component, with the excess already starting near the sublimation temperature. This is less than about 14 AU from the central star in all objects.
- The energy balance calculations indicate a non-spherical distribution of the circumstellar dust for 5 sources. For R Sct, however, we present evidence that the distance based on the $P - L$ relation is wrong and R Sct must be significantly less luminous.
- At far-IR and millimetre wavelengths we are sampling a black-body slope in the flux distribution. For all stars, except R Sct, the spectral index $p$ lies in the range 0.0–0.4. The IRAS-submillimetre flux distributions are thus consistent with emission by large ($\gtrsim 0.1$ mm) grains.

Direct evidence for present day mass-loss is absent, and we infer from our findings that at least part of the circumstellar dust must be stored in a long-lived structure, likely a disc. First-order estimates imply that the discs are small and, given the large distance of the sample, they are likely only resolvable with interferometric observations. We speculate that the discs are essentially Keplerian since any outflow velocity would bring the dust much further from the star in the timescale of the central stars' evolution from the AGB to the RV Tau instability strip. As the redistribution of star light is very efficient, the height above the midplane of those discs must be large.

The most obvious way to explain the formation of a disc around evolved RV Tauri stars is to assume that the disc was created during binary interaction, when the primary star was at giant dimensions. Binarity in RV Tauri stars, however, is hard to prove with radial velocity monitoring since the photospheric pulsations have large amplitudes in radial velocity. Only in a few examples was orbital motion indeed found: AC Her (Van Winckel et al. 1998), U Mon (Pollard & Cottrell 1995) and RU Cen and SX Cen (Maas et al. 2002). In this analysis we confirm the proposition made earlier (Van Winckel et al. 1999) that the six well-studied dusty RV Tauri stars are likely all binaries in which the binary interaction has played a fundamental role in creating a stable dusty disc.

The actual structure of the disc, as well as its formation, stability and evolution are not well understood at this stage. We argue that realistic models of circumstellar material of RV Tauri stars should be constructed in the framework of stable discs, similar to the post-AGB star HR 4049 (Dominik et al. 2003).

*Acknowledgements.* The staff and service observers of the Mercator Observatory at La Palma are acknowledged for the photometric Geneva data. This publication makes also use of data from the Two Micron All Sky Survey, which is a joint project of the University of Massachusetts and the Infrared Processing and Analysis Center/California Institute of Technology, funded by the National Aeronautics and Space Administration and the National Science Foundation. We also used data from the DENIS project, which is partly funded by the European Commission through SCIENCE and Human Capital and Mobility grants. It is also supported in France by INSU, the Education Ministry and CNRS, in Germany by the Land of Baden-Würtenberg, in Spain by DGICYT, in Italy by CNR, in Austria by the Fonds zur Förderung der Wissenschaftlichen Forschung and Bundesministerium für Wissenschaft und Forschung. We also like to thank the anonymous referee for his contribution to this paper.

# Online Material

## Appendix A: Photometric data

**Table A.1.** Geneva data were acquired with the Flemish Mercator Telescope at La Palma, Spain. Our total dataset was scanned for the maximum and minimum magnitudes. Observation dates, number of measurements and total timebases of these maxima and minima are given as well. For AC Her, our observational monitoring data are divided in two major blocks of 8 and 446 days with 4612 days inbetween. Additional data were found in the Geneva database (http://obswww.unige.ch/gcpd/gcpd.html).

| Name | JD | Number of measurements | Total time base (days) | U | B | V | B1 | B2 | V1 | G | Reference |
|---|---|---|---|---|---|---|---|---|---|---|---|
| RV Tau | 2 452 305.412 | 53 | 834 | 12.511 | 9.957 | 9.036 | 11.323 | 11.077 | 9.859 | 9.886 | Mercator |
|  | 2 452 609.487 |  |  | 14.867 | 12.091 | 10.822 | 13.507 | 13.130 | 11.650 | 11.655 | Mercator |
| SU Gem | 2 452 543.715 | 29 | 783 | 14.301 | 11.616 | 10.697 | 12.902 | 12.808 | 11.539 | 11.540 | Mercator |
|  | 2 452 307.508 |  |  | 16.610 | 14.028 | 13.248 | 15.160 | 15.386 | 14.042 | 14.057 | Mercator |
| U Mon |  |  |  | 8.428 | 6.330 | 5.980 | 7.596 | 7.508 | 6.752 | 6.948 | GCPD |
|  | 2 447 545.766 | 119 | 1217 | 7.759 | 5.792 | 5.575 | 7.013 | 6.990 | 6.334 | 6.570 | Mercator |
|  | 2 447 853.822 |  |  | 9.507 | 7.484 | 7.194 | 8.697 | 8.698 | 7.941 | 8.245 | Mercator |
| AC Her |  |  |  | 9.419 | 7.450 | 7.456 | 8.536 | 8.755 | 8.221 | 8.460 | GCPD |
|  | 2 452 367.729 | 62 | 5039 | 8.726 | 6.900 | 7.261 | 7.854 | 8.311 | 8.002 | 8.310 | Mercator |
|  | 2 452 097.548 |  |  | 10.770 | 8.729 | 8.408 | 9.927 | 9.941 | 9.187 | 9.354 | Mercator |
| R Sct |  |  |  | 9.073 | 6.316 | 5.504 | 7.720 | 7.415 | 6.306 | 6.406 | GCPD |
|  | 2 447 756.567 | 36 | 3406 | 7.875 | 5.416 | 4.824 | 6.735 | 6.561 | 5.621 | 5.727 | Mercator |
|  | 2 447 793.506 |  |  | 10.165 | 7.341 | 6.614 | 8.689 | 8.477 | 7.399 | 7.577 | Mercator |

**Table A.2.** Ground-based optical data, acquired over a long period, found in the literature.

| Name | U | B | V | R | I | System | Reference |
|---|---|---|---|---|---|---|---|
| TW Cam | 12.12 | 10.94 | 9.51 |  |  | Johnson | Dawson (1979) |
| RV Tau | 12.52 | 10.93 | 9.19 |  |  | Johnson | Dawson (1979) |
| SU Gem | 12.88 | 11.70 | 10.19 |  |  | Johnson | Dawson (1979) |
| UY CMa | 11.36 | 11.03 | 10.4 |  |  | Johnson | Dawson (1979) |
|  | 12.28 | 11.81 | 11.0 |  |  | Johnson | Dawson (1979) |
| U Mon | 7.15 | 6.59 | 5.66 |  |  | Johnson | Dawson (1979) |
|  |  | 6.497 | 5.436 | 4.946 | 4.511 | Cousins | Pollard et al. (1996) |
|  |  | 8.614 | 7.632 | 7.088 | 6.387 | Cousins | Pollard et al. (1996) |
| AC Her | 8.00 | 7.66 | 7.03 |  |  | Johnson | Dawson (1979) |
|  | 9.11 | 8.68 | 7.87 | 7.40 | 6.91 | Cousins | Goldsmith et al. (1987) |
| R Sct | 8.00 | 6.52 | 5.08 |  |  | Johnson | Dawson (1979) |
|  |  | 5.999 | 4.807 | 4.103 | 3.589 | Cousins | Pollard et al. (1996) |
|  |  | 9.368 | 7.769 | 7.004 | 5.453 | Cousins | Pollard et al. (1996) |

**Table A.3.** Ground-based near-IR photometry found in the literature. If there was more than one measurement we used both the maximum and minimum data points. For the data from 2 MASS and DENIS we made use of the catalogues found in VIZIER (http://vizier.u-strasbg.fr/viz-bin/VizieR).

| Name | I | J | H | K | L | M | System | Reference |
|---|---|---|---|---|---|---|---|---|
| TW Cam | | | | | 4.2 | 3.5 | Mt Lemmon, Arizona | Gehrz (1972) |
| | | 7.04 | 6.36 | 5.70 | 4.40 | | SAAO | Lloyd Evans (1985) |
| | | 7.035 | 6.364 | 5.750 | | | 2 MASS | VIZIER |
| RV Tau | | | | 5.1 | 3.3 | 2.3 | Mt Lemmon, Arizona | Gehrz (1972) |
| | | 6.39 | 5.67 | 5.03 | 3.64 | | SAAO | Lloyd Evans (1985) |
| | | 6.183 | 5.488 | 4.777 | | | 2 MASS | VIZIER |
| SU Gem | | | | | 4.1 | 3.2 | Mt Lemmon, Arizona | Gehrz (1972) |
| | | 8.36 | 7.38 | 6.47 | 4.94 | | SAAO | Lloyd Evans (1985) |
| | | 8.97 | 8.38 | 6.95 | 4.96 | | SAAO | Lloyd Evans (1985) |
| | | 10.167 | 9.033 | 7.584 | | | 2 MASS | VIZIER |
| UY CMa | | 9.35 | 8.82 | 8.22 | 6.62 | | SAAO | Lloyd Evans (1985) |
| | | 9.42 | 8.46 | 8.41 | 6.87 | | SAAO | Lloyd Evans (1985) |
| | 9.845 | 9.093 | | 8.038 | | | DENIS | VIZIER |
| | | 9.042 | 8.597 | 7.986 | | | 2 MASS | VIZIER |
| U Mon | | | | | 2.4 | 1.5 | Kitt Peak NO, Arizona | Gehrz & Woolf (1970) |
| | | | | 3.7 | 2.7 | 1.7 | Mt Lemmon, Arizona | Gehrz (1972) |
| | | 4.01 | 3.65 | 3.25 | 2.40 | | SAAO | Lloyd Evans (1985) |
| | | 4.88 | 4.43 | 4.08 | 3.08 | | SAAO | Lloyd Evans (1985) |
| | | 4.35 | 3.93 | 3.60 | 2.30 | 1.61 | ESO | Bogaert (1994) |
| | | 4.925 | 4.269 | 4.042 | | | 2 MASS | VIZIER |
| AC Her | | | | | 4.4 | 4.0 | Kitt Peak NO, Arizona | Gehrz & Woolf (1970) |
| | | | | 5.2 | 4.7 | 3.6 | Mt Lemmon, Arizona | Gehrz (1972) |
| | | 5.97 | 5.52 | 5.28 | 4.73 | | SAAO | Lloyd Evans (1985) |
| | | 6.21 | 5.66 | 5.40 | 4.87 | 4.55 | SAAO | Goldsmith et al. (1987) |
| | | 5.700 | 5.338 | 5.075 | | | 2 MASS | VIZIER |
| R Sct | | | | | 1.6 | 1.3 | Kitt Peak NO, Arizona | Gehrz & Woolf (1970) |
| | | | | 2.8 | 1.8 | 1.4 | Mt Lemmon, Arizona | Gehrz (1972) |
| | | 2.78 | 2.28 | 2.02 | 1.51 | | SAAO | Lloyd Evans (1985) |
| | | 3.38 | 3.04 | 2.80 | 2.33 | | SAAO | Lloyd Evans (1985) |
| | | 2.84 | 2.26 | 2.00 | 1.49 | | SAAO | Goldsmith et al. (1987) |
| | | 2.87 | 2.27 | 2.01 | 1.53 | | SAAO | Goldsmith et al. (1987) |
| | | 2.824 | 2.374 | 2.156 | | | 2 MASS | VIZIER |

**Table A.4.** IRAS photometry points at 12, 25, 60 and 850 $\mu$m. Note however that in some cases the 100 $\mu$m is an upper limit (L); these observations are probably contaminated by interstellar cirrus clouds. For RV Tau the 100 $\mu$m data point is a lower limit (:).

| Name | $F_{12}$(Jy) | $F_{25}$(Jy) | $F_{60}$(Jy) | $F_{100}$(Jy) |
|---|---|---|---|---|
| TW Cam | 8.25 | 5.60 | 1.84 | 1.79L |
| RV Tau | 22.52 | 18.05 | 6.50 | 2.44: |
| SU Gem | 7.91 | 5.68 | 2.19 | 11.99L |
| UY CMa | 3.49 | 2.46 | 0.57 | 1.00L |
| U Mon | 124.30 | 88.43 | 26.59 | 9.54 |
| AC Her | 41.43 | 65.33 | 21.37 | 8.04 |
| R Sct | 20.83 | 9.26 | 8.20 | 140.30L |

**Table A.5.** Data of the Midcourse Space eXperiment (MSX). The instrument on board MSX is the SPIRIT III (Spatial Infrared Imaging Telescope III). The approximate effective wavelengths of the 6 MSX filters are in $\mu$m.

| Name | $B1$(Jy) 4.29 $\mu$m | $B2$(Jy) 4.35 $\mu$m | $A$(Jy) 8.28 $\mu$m | $C$(Jy) 12.13 $\mu$m | $D$(Jy) 14.65 $\mu$m | $E$(Jy) 21.34 $\mu$m |
|---|---|---|---|---|---|---|
| SU Gem | | | 8.283 | 7.666 | 7.178 | 7.505 |
| U Mon | | 42.93 | 54.45 | 88.71 | 72.38 | 73.29 |
| R Sct | 43.12 | 49.27 | 22.02 | 11.82 | 11.09 | 6.350 |